\documentclass[twocolumn,english,aps,prl]{revtex4}
\usepackage[T1]{fontenc}
\usepackage[latin1]{inputenc}
\usepackage{graphicx}
\usepackage{amssymb}
\makeatletter
\usepackage{babel}
\makeatother

\begin{document}
\title{Cauchy-Schwarz characterization of tripartite \\ quantum correlations 
in an optical parametric oscillator}

\author{K. Dechoum, W. S. Marques, and  A. Z. Khoury}

\affiliation{Instituto de F\'{\i}sica - Universidade Federal Fluminense\\
24210-346, Niter\'{o}i-RJ, Brazil}

\begin{abstract}

We analyze the three-mode correlation properties of the electromagnetic field in a optical 
parametric oscillator below threshold. We employ a perturbative expansion of the It\^o 
equations derived from the positive-P representation of the density matrix. Using the 
generalized Cauchy-Schwarz inequality, we investigate the genuine quantum nature of the 
triple correlations between the interacting fields, since in this case continuous variable 
entanglement is not detected by the van Loock-Furusawa criterion 
[Phys. Rev. A {\bf 67}, 052315 (2003)]. 
Although not being a necessary condition, these triple correlations are a sufficient 
evidence of tripartite entanglement. Of course, our characterization of the 
quantum correlations is applicable to non-Gaussian states, which we show to 
be the case of the optical parametric oscillator below threshold, 
provided nonlinear quantum fluctuations are properly taken into account.

\end{abstract}
\maketitle

The nonlocal character of quantum mechanics has been widely exploited to develop quantum 
information protocols~\cite{protocolsIQ} sometimes implemented in laboratories~\cite{protocolsIQlab}, 
or even for commercial purposes as has occurred with quantum cryptography~\cite{commQcript}. 
Bipartite entangled states are the simplest resource for such protocols and can be generated by 
optical parametric oscillators (OPOs)
\cite{nossojosab,nossowigner,hyper,phaseconj}. 
However, more sophisticated protocols may require entanglement of many parts~\cite{OPOprotocols1,OPOprotocols2}. 
The next difficulty is trying to find a source of tripartite entanglement. The first idea in this sense 
came from Greenberger, Horne, and Zeilinger~\cite{GHZ}, suggesting the GHZ state, a discrete variable 
(spin variables) tripartite entangled state detectable by triple product spin measurements, but actually very 
difficult to implement. However the implementation of this idea by using continuous variables such as  
quadratures of optical fields, seems quite feasible with the available homodyning techniques.
Nowadays, there are a few devices that can generate multipartite entangled states, and 
the OPO has been explored for this task \cite{science-usp}.

The nonlinear interaction between the OPO modes suggests the possibility of  
tripartite entanglement between the three modes, and also the possibility of controlling the
bipartite correlations between signal and idler only by handling a third part, the pump mode.
The quantum correlations between the different parts are supposed to change depending on
statistics, spatial profile and other possible characteristics of the pump input, as observed 
in many experiments based on parametric down conversion~\cite{mineirosetal}.
Indeed, tripartite pump-signal-idler entanglement has been predicted in the intense regime, 
above the OPO threshold \cite{nussenz1} and experimentally generated \cite{nussenz2}.
Tripartite correlations were analyzed according to the criteria proposed by van Loock and Furusawa\cite{vanloock}, 
based on pair correlations among the three modes. 
Also, there are other experiments that generate continuous-variable tripartite entangled states
obtained by mixing squeezed vacua with linear optical elements \cite{tripartite1}.

However, there has been little attention to tripartite correlations in the OPO 
operating below threshold, but it was in this context that triple 
quadrature correlations were proposed to be measured as means to compare classical and quantum 
correlations~\cite{chaturvedi}. The experimental difficulty resides in 
homodyning the weak down converted beams, which is not a problem for intense beams generated 
above threshold, since a smart technique of self-homodyning \cite{nussenz2} can be used.
It is also important to notice that it was in the below threshold regime where squeezed vacuum was first 
produced, and EPR correlations were first detected \cite{silvania1,silvania2}. 

In this work, we calculate, using a perturbative approach, all triple quadrature correlations 
between the interacting fields in the OPO \textbf{below} threshold. These triple correlations 
should vanish for Gaussian quantum states, whether they are entangled or not, so that the 
van Loock and Furusawa\cite{vanloock} entanglement criterion is well suited. This is the 
case, for example, of the OPO \textbf{above} threshold \cite{science-usp}. 
However, we shall demonstrate that triple 
quadrature correlations are predicted for below threshold operation when the perturbative 
approach is taken beyond the linear approximation. While evidencing non Gaussian behavior, 
these triple correlations are also sufficient conditions for tripartite entanglement. 
We also show that, in this case, the van Loock-Furusawa criterion fails to evidence entanglement. 
Due to this inconclusive condition, and the lack of a criterion for this case, we also show that 
certain correlations violate the tripartite Cauchy-Schwarz inequality expected for classical 
c-number variables. 

The OPO can be modeled by three modes coupled through a nonlinear crystal 
inside a lossy Fabry-Perot cavity, described by the following master equation 
in the Born-Markov approximation \cite{carmichael}: 
\begin{eqnarray}
\frac{\partial \hat{\rho}}{\partial t} = 
-\frac{i}{\hbar} \left[ \hat{H}, \hat{\rho} \right] 
+\sum_{i=0}^{2} \gamma_{i} \left( 2 \hat{a}_{i} \hat{\rho} 
\hat{a}_{i}^{\dagger} - \hat{a}_{i}^{\dagger} \hat{a}_{i} \hat{\rho} -
\hat{\rho} \hat{a}_{i}^{\dagger} \hat{a}_{i} \right)\,,
\label{a2}
\end{eqnarray}
where 
\begin{eqnarray}
\hat{H} &=& \sum_{i=0}^{2} \hbar \omega_{i} \hat{a}_{i}^{\dagger} \hat{a}_{i}
+ i \hbar \chi \left(\hat{a}_{1}^{\dagger} \hat{a}_{2}^{\dagger}
\hat{a}_{0} - \hat{a}_{1} \hat{a}_{2} \hat{a}_{0}^{\dagger} \right)
\label{a1}\\ 
&+&
i\hbar \left( E e^{-i\omega_{0} t}\hat{a}_{0}^{\dagger} -
E^{*} e^{i\omega_{0} t}\hat{a}_{0} \right)\,.
\nonumber 
\end{eqnarray}
The parameter $E$ represents the external pump field at frequency 
$\omega_{0}$. The pump, signal and idler fields are represented by 
the annihilation operators $\hat{a}_{0}$, $\hat{a}_{1}$ and $\hat{a}_{2}$, respectively, 
and satisfy the frequency match condition $\omega_{0}=\omega_{1}+\omega_{2}$. 
$\chi$ is the nonlinear coupling and $\gamma_{i}$ ($i=0,1,2$) are 
amplitude damping rates. 
Now, the density matrix equation of motion can be transformed into c-number 
Fokker-Planck or stochastic equations by means of operator representation theory.

The quantum evolution of the system can be described by the generalized 
P-representation~\cite{representacaoP}, by expanding the density matrix in a off 
diagonal coherent state basis, defined as
\begin{equation}
\hat{\rho} = \int_{\cal{D}} \frac{| \alpha \rangle \langle 
\left( \alpha^{+}\right)^{*} |}
{\langle \left( \alpha^{+} \right)^{*} | \alpha \rangle} 
P(\alpha,\alpha^{+})d^{6}\alpha d^{6}\alpha^{+}\,,
\label{a3}
\end{equation}
where $\alpha \equiv \left(\alpha_{0},\alpha_{1},\alpha_{2}\right)$ and 
$\alpha^{+} \equiv \left(\alpha_{0}^{+},\alpha_{1}^{+},\alpha_{2}^{+}\right)$
are complex variables that run independently over all complex plane.
The function $P(\alpha,\alpha^{+})$ can be understood as a positive 
distribution in a double phase space that satisfies a Fokker-Planck equation.
The It\^{o} stochastic equations derived from the Fokker-Planck equation,
after transforming to the rotating-wave frame, are
\begin{eqnarray}
d \alpha_{0} &=& \left(E - \gamma_{0} \alpha_{0} - \chi 
\alpha_{1} \alpha_{2} \right) dt \nonumber \\
d \alpha_{0}^{+} &=& \left(E^{*} -\gamma_{0} \alpha_{0}^{+} 
- \chi \alpha_{1}^{+} \alpha_{2}^{+} \right) dt \nonumber \\
d \alpha_{j} &=& \left(-\gamma_{j} \alpha_{j} + \chi 
\alpha_{k}^{+} \alpha_{0} \right) dt 
+ \left( \chi \alpha_{0} \right)^{1/2} d W_{j} \nonumber \\
d \alpha_{j}^{+} &=& \left(-\gamma_{j} \alpha_{j}^{+}+ \chi 
\alpha_{k} \alpha_{0}^{+} \right) dt + \left( \chi 
\alpha_{0}^{+} \right)^{1/2} d W_{j}^{+}\,,
\label{a6}
\end{eqnarray}
where $j=1,2$; $k=1,2$ and $j\neq k$. The Wiener increments satisfy 
$\langle d W_{1} d W_{2} \rangle = \langle d W_{1}^{+} 
d W_{2}^{+} \rangle = dt\,,
$
%
and all other correlations vanish.

In order to analyze the correlations between the modes, we proceed by applying the 
perturbation theory for the three modes below threshold, where this approach is valid. Around 
the threshold, the quantum fluctuations become huge and break the perturbative analysis.
Following the physical situation found in most experiments, we shall assume a common 
damping rate for the down converted fields: $\gamma_1=\gamma_2=\gamma$. It is also 
useful to define a dimensionless coupling constant $g=\chi/(\gamma\sqrt{2\gamma_r})$, 
where $\gamma_r\equiv\gamma_0/\gamma$. In order to employ a perturbative approach and 
to decouple the dynamical equations, it is convenient to define the following 
normalized variables \cite{caves}: 
\begin{eqnarray}
x_{0}&=& g \sqrt{2 \gamma_{r}} \left ( \alpha_{0} + \alpha_{0}^{+} \right ) 
=  g \sqrt{2 \gamma_{r}} X_{0} \nonumber \\
y_{0}&=& g \sqrt{2 \gamma_{r}}\, \frac{1}{i} \left( \alpha_{0} - \alpha_{0}^{+} \right ) 
=  g \sqrt{2 \gamma_{r}} Y_{0}  \nonumber \\
x&=& g \left ( \alpha_{1} + \alpha_{2}^{+} \right ) = g X \nonumber \\
y&=& g \frac{1}{i}\, \left( \alpha_{1} - \alpha_{2}^{+} \right ) = g Y \;.
\end{eqnarray}
Note that in positive-P representation the quadrature variables $x$ and $y$ are not 
real and their definitions are accompanied by the corresponding phase space conjugates 
$x^{+}$ and $y^{+}$. 
We now expand these variables in terms of the dimensionless coupling constant \cite{kaled2004A,kaled2004B}: 
\begin{equation}
x_{0} = \sum_{n=0}^{\infty} g^n x_{0}^{(n)} = x_{0}^{(0)} + g  x_{0}^{(1)} + g^{2}  x_{0}^{(2)} + ...\;, 
\end{equation}
and analogous expansions for $y_{0}, x, y, x^{+}$ and $y^{+}$. 

Below threshold, the zero-order steady state solution for the quadratures are given by
\begin{eqnarray}
x^{(0)} &=& {x^{+}}^{(0)} = y^{(0)} = {y^{+}}^{(0)} = y_{0}^{(0)} = 0 
\nonumber\\
x_{0}^{(0)} &=& 2 \mu\;, 
\end{eqnarray}
where $\mu=\chi E/(\gamma\gamma_0)$ is the dimensionless pump parameter. 
These are the mean values of the fields below threshold. As expected, at this regime all mean field quadratures  
are zero, since only fluorescence is present. The pumped quadrature is the only one having a macroscopic value. 
The first quantum correction comes from the next order contribution:
\begin{eqnarray}
x^{(1)} (t) &=& \sqrt {2 \mu} \int_{-\infty}^{t} e^{-(1-\mu)(t-t')} \xi_{x}(t') dt' \nonumber \\
y^{(1)} (t) &=& -i \sqrt {2 \mu} \int_{-\infty}^{t} e^{-(1+\mu)(t-t')} \xi_{y}(t') dt' \nonumber \\
x_{0}^{(2)} (t) &=& - \gamma_{r} \int_{-\infty}^{t} e^{-\gamma_{r}(t-t')} \left (
x^{(1)} (t') {x^{+}}^{(1)} (t') \right. \nonumber \\ 
&-& \left. y^{(1)} (t') {y^{+}}^{(1)} (t') \right ) dt' \nonumber \\
y_{0}^{(2)} (t) &=& - \gamma_{r} \int_{-\infty}^{t} e^{-\gamma_{r}(t-t')} \left (
x^{(1)} (t') {y^{+}}^{(1)} (t') \right. \nonumber \\ 
&+& \left. y^{(1)} (t') {x^{+}}^{(1)} (t') \right ) dt' \;,
\end{eqnarray}
where,
$\langle \xi_{x}(t)\xi_{x^{+}}(t^ {\prime})\rangle = 
\langle \xi_{y}(t)\xi_{y^{+}}(t^ {\prime}) \rangle = \delta ( t - t')\;.
$
Note that we kept the second order perturbation term for the pumped mode variables to 
account for pump depletion, since their first order corrections vanish. Therefore, pump 
depletion is a second order effect below threshold. 
It is also interesting to note that the second order corrections in the quadrature variables of the pumped mode 
depend on the product of two correlated Gaussian processes. This implies that double correlations between 
the pump and any one of the down converted modes vanish. As a result, a criterion based on pair correlations 
for tripartite entanglement, like the one by van Loock and Furusawa \cite{vanloock}, fails to detect the existing 
triple correlations which are quantum mechanical in essence, as we shall see shortly. Moreover, the second order 
correction to the pump fluctuations are not Gaussian, since it can be checked that its odd moments do not 
vanish. 

With the presented solutions, we can calculate any kind of correlations between the mode variables.
Correlations between two quadratures generate the expected 
two-mode squeezed state for $yy^{+}$. Actually, the four variables of the down converted modes 
can be combined to form EPR variables. An important observation is that there are no two-mode correlations 
between any variable of the pumped mode and the down converted ones, since the pumped mode depend on the product 
of two Gaussian noises. Due to this property, the three-mode correlations unveil most of the important result 
addressed by this paper.

The results for the steady state time ordered triple correlations calculated in normal order, 
is presented below. Actually these results are also the output cavity fields, since in the 
positive-P representation the incoming field (vacuum) is not correlated with the intracavity modes.
The non vanishing triple correlations up to fourth order in the coupling constant are:
\begin{eqnarray}
\langle \Delta x\,\Delta x^{+}\Delta x_{0} \rangle &=& - g^{4} \left ( \frac{\mu}{1 - \mu} \right )^{2} 
\left[  \frac{2}{1 + \mu} \right. \nonumber \\
&+&\left. \frac {\gamma_{r}}{\gamma_{r} + 2 \left(1-\mu \right)} \right ] \nonumber \\
\langle \Delta y\,\Delta y^{+} \Delta x_{0} \rangle &=& g^{4} \left ( \frac{\mu}{1 + \mu} \right )^{2} 
\left[ \frac{2}{1 - \mu} \right. \nonumber \\
&+& \left.\frac {\gamma_{r}}{\gamma_{r} + 2 \left(1+\mu \right)} \right ] \nonumber \\
\langle \Delta y\,\Delta x^{+}\Delta y_{0} \rangle &=& g^{4} \left ( \frac{\mu^{2}}{1 - \mu^{2}} \right ) 
\left ( \frac{\gamma_{r}}{2 + \gamma_{r}} \right )\nonumber \\
\langle \Delta x\,\Delta y^{+} \Delta y_{0} \rangle &=& g^{4} \left ( \frac{\mu^{2}}{1 - \mu^{2}} \right ) 
\left ( \frac{\gamma_{r}}{2 + \gamma_{r}} \right )\;,
\label{triple}
\end{eqnarray}
where $\Delta u\equiv u-\langle u\rangle$. One can easily show that for any kind of separable 
state of the form: $\sum p_i \rho_k\otimes\rho_{lm}$, where $\{k,l,m\}$ is any permutation of 
$\{0,1,2\}$, at least one of these triple correlations should vanish identically. Therefore, 
the set of triple correlations given by eq.(\ref{triple}) is a sufficient condition for 
tripartite entanglement. However, it does not exclude classical correlations \cite{luciano},  
so that further analysis is required to evidence genuine quantum behavior. 

Let us now turn to a reliable classicality criterion based on Cauchy-Schwarz inequalities. 
First, we calculate the average product of the down converted squared fluctuations: 
\begin{eqnarray}
\langle (\Delta x^{2} &+& \Delta y^{2})\,[(\Delta x^{+})^{2} + (\Delta {y^{+}})^{2}] \rangle = 
\nonumber\\
&&2\,g^{4}\,\mu^{2} \left( \frac{1}{\left(1 - \mu \right)^{2}} +\frac{1}{\left(1 + \mu \right)^{2}} \right)\;.
\label{deltaII}
\end{eqnarray}
The fluctuations on $x_{0}$ are given by
\begin{eqnarray}
\langle \left( \Delta x_{0} \right)^{2} \rangle &=&  
g^{4} \left( \frac{\mu}{1 - \mu} \right)^2 
\left\{ \left( \frac{2}{1 + \mu} \right)^2 \right. 
\label{deltax0}\\ 
&+& \left. \left[ 1 + \left(\frac{1-\mu}{1+\mu} \right)^2  \right]
\frac{\gamma_{r}^{2}}{\gamma_{r}^{2} + 4 \left(1-\mu \right)^{2}}   \right\} \;,
\nonumber
\end{eqnarray}
while $\langle \left( \Delta y_{0} \right)^{2}\rangle$ is negligible. 
Now we establish a relation between correlations (\ref{triple}) and intensity correlations.
To begin with, we assume that for any three complex numbers the following relation holds
\begin{equation}
\langle | \alpha_{1} \alpha_{2} + \lambda^{*} \alpha_{0}^{*} |^{2} \rangle \ge 0 \;,
\end{equation}
so that
\begin{equation}
\langle\alpha_{1}^{*} \alpha_{1} \alpha_{2}^{*} \alpha_{2} \rangle + |\lambda|^{2} 
\langle \alpha_{0}^{*} \alpha_{0} \rangle + \lambda \langle \alpha_{1}\alpha_{2}\alpha_{0}\rangle + 
\lambda^{*} \langle \alpha_{1}^{*} \alpha_{2}^{*} \alpha_{0}^{*}\rangle \ge 0\;.
\end{equation}
This expression is minimized for
$\lambda = \langle \alpha_{1}^{*} \alpha_{2}^{*} \alpha_{0}^{*}\rangle/\langle\alpha_{0}^{*}\alpha_{0} 
\rangle$, 
%
giving rise to the following inequality:
\begin{equation}
\langle\alpha_{1}^{*} \alpha_{1} \alpha_{2}^{*} \alpha_{2} \rangle \langle\alpha_{0}^{*}\alpha_{0} \rangle
\ge |\langle \alpha_{1}\alpha_{2}\alpha_{0}\rangle |^{2}\,.
\label{schwartz}
\end{equation}
In the same way, similar inequalities can also be derived by permutation of the indices in the 
complex variables $\alpha_j$. After a straightforward algebra, it is easy to show that the 
triple correlation appearing in the right hand side of inequality (\ref{schwartz}) is simply 
the squared sum of all triple correlations given by eq.(\ref{triple}), taking the first one 
($\langle \Delta x\,\Delta x^{+}\Delta x_{0} \rangle$) with an inverted sign. Also, the left 
hand side of the inequality is simply the product of the averages given by eqs.(\ref{deltaII}) 
and (\ref{deltax0}). 

Using the results given by eqs.(\ref{triple}), (\ref{deltaII}), and (\ref{deltax0}) 
on both sides of the Cauchy-Schwarz inequality (\ref{schwartz}), we can verify its 
violation for the three-mode state produced by the OPO operating below threshold, which means that these triple correlations are essentially quantum mechanical. The reliability of the analytical calculations can be checked by numerical simulation of the full set of stochastic equations (\ref{a6}), \textbf{without any perturbative approximation}. 
Fig.(1) compares both sides of inequality (\ref{schwartz}), showing a clear violation of the inequality for large $\gamma_r$. The numerical results are in very good agreement with the analytical calculations. In our numerical calculations we used $g=0.0071$ ($\chi=0.01$) whose order of magnitude corresponds to most nonlinear media used for parametric down conversion. Note that the violation of Cauchy-Schwarz inequality occurs with very small correlation values ($\sim 10^{-4}$). This is due to the rather small value of the coupling constant. Since these correlations scale as $g^2$, this result can be substantially improved with new materials providing sgnificant values of the nonlinear susceptibility. Indeed, we checked this numerically. 

In conclusion, we investigated the triple correlations produced in an OPO below 
threshold, evidencing tripartite entanglement not detected using the 
well known van Loock-Furusawa criterion. 
In order to circumvent this limitation, we employed a non classicality criterion based on 
general Cauchy-Schwartz inequalities for third order moments. This approach was applicable 
due to the non Gaussian character of the quantum fluctuations. Moreover, these triple 
correlations rule out separable states, being therefore a sufficient 
condition for entanglement. 
Although these inequalities have already been used for discrete 
variables \cite{GHZ}, this is the first time, to our knowledge, they are used to characterize 
non classical triple correlations in continuous variables. 
Whether these triple correlations 
can be useful for quantum information protocols remains to be investigated. 

\begin{figure}
\resizebox*{7.7cm}{5.5cm}{\rotatebox{-90}{\includegraphics{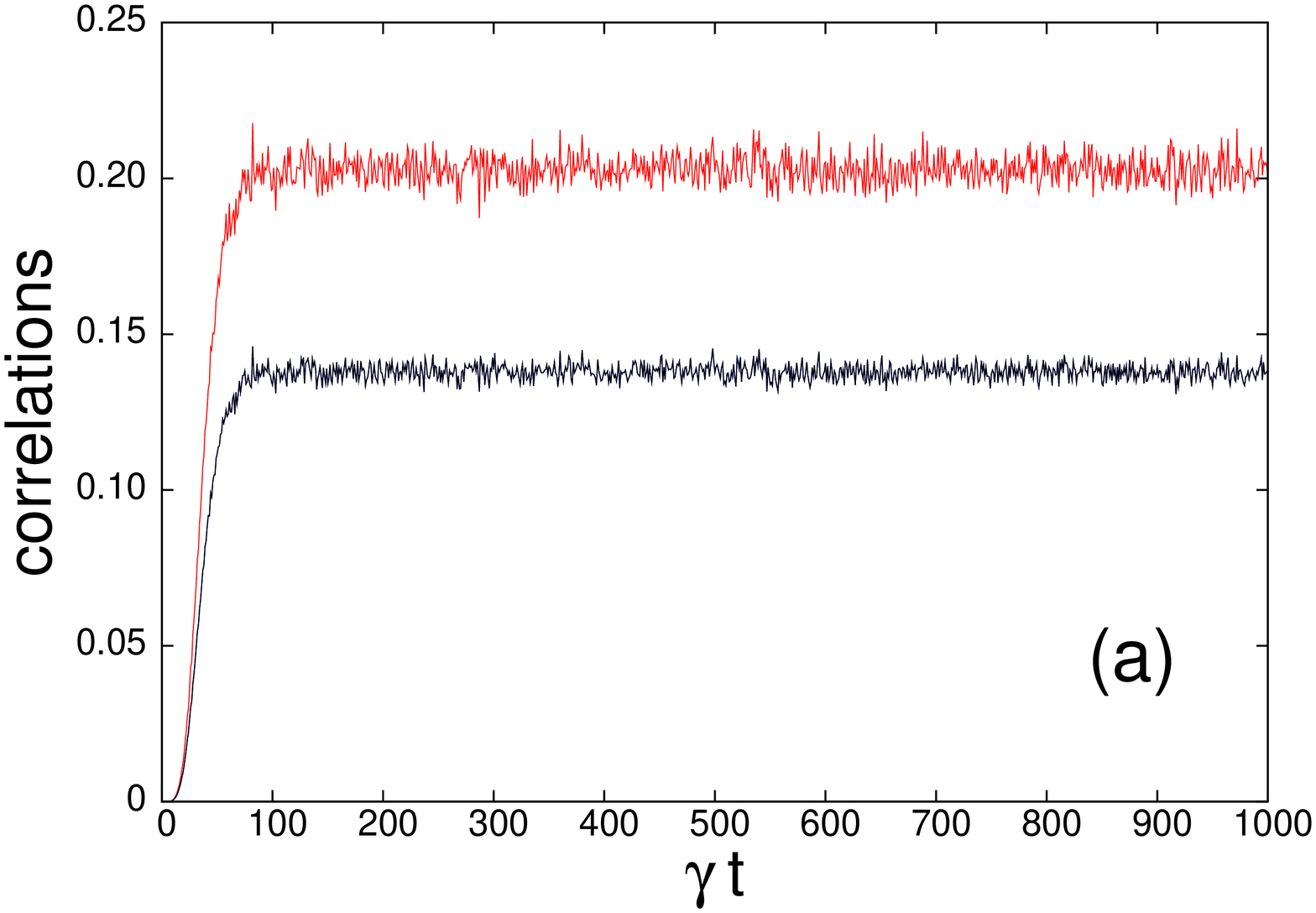}}}
\resizebox*{7.7cm}{5.5cm}{\rotatebox{-90}{\includegraphics{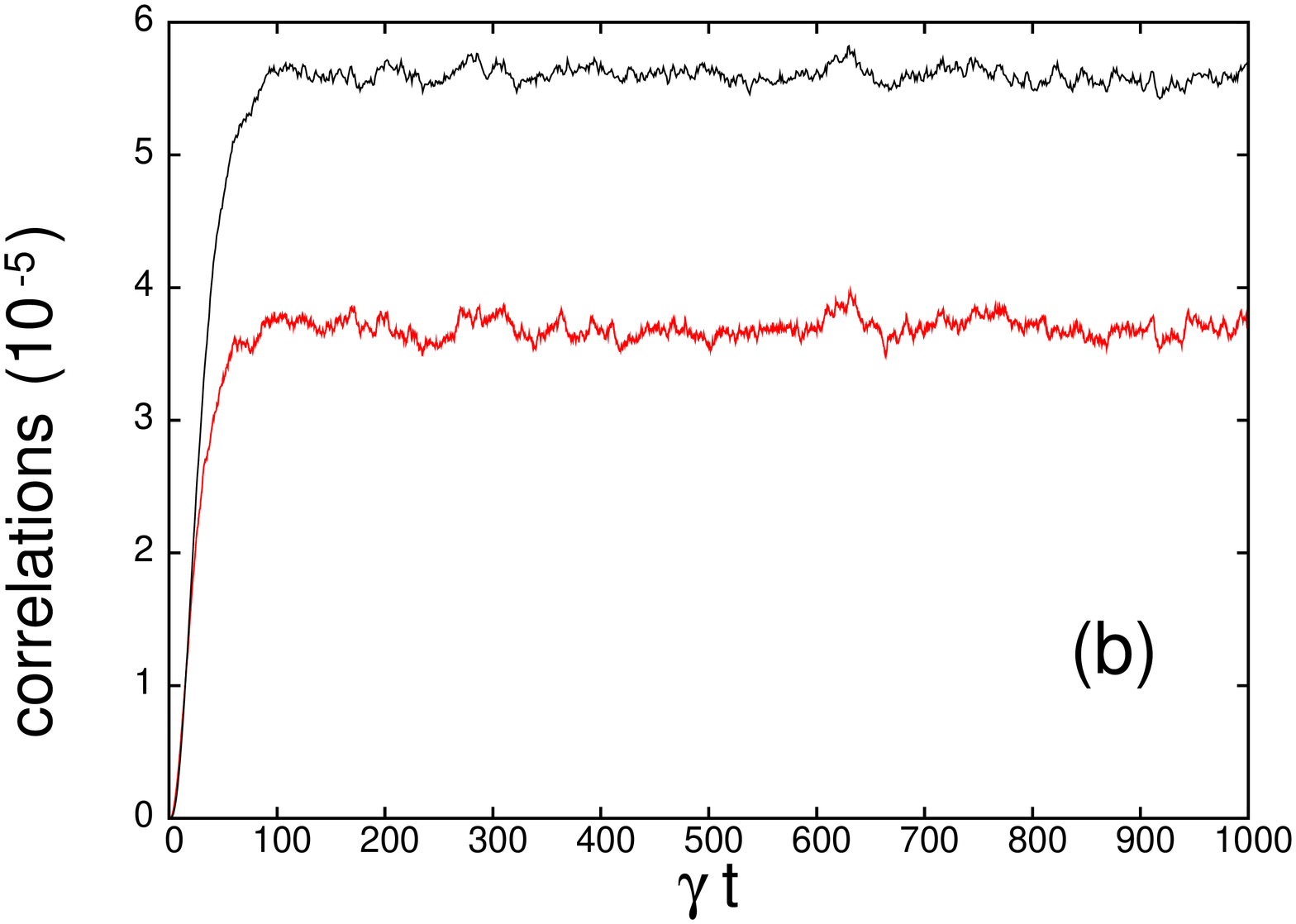}}}
\caption{Violation of Cauchy-Schwarz inequality for an OPO operating below 
threshold. The figures show simulations using $\mu = 0.7$ and $g=0.0071$. 
(a) Bottom curve (black) represents the right hand side of inequality 
(\ref{schwartz}), and top curve (red online) represents the left hand side 
for $\gamma_{r} = 0.01$. The Cauchy-Schwarz inequality is not violated. 
(b) Top curve (black) represents the right hand side of inequality 
(\ref{schwartz}), and bottom curve (red online) represents the left hand side 
for $\gamma_{r} = 100$. Violation of the Cauchy-Schwarz inequality is clear.} 
\end{figure}

\begin{acknowledgments}

This work was supported by the Instituto Nacional de 
Ci\^encia e Tecnologia de 
Informa\c c\~ao Qu\^antica (INCT/IQ - CNPq - Brazil), 
Coordena\c c\~ao de 
Aperfei\c coamento de Pessoal de N\'{\i}vel Superior (CAPES - Brazil) and Funda\c c\~ao 
de Amparo \`a Pesquisa do Estado do Rio de Janeiro (FAPERJ).

\end{acknowledgments}

\end{document}